\documentclass[aps,prl,twocolumn]{revtex4}

\usepackage{graphicx,latexsym,color}

\def\be{\begin{equation}}

\def\ee{\end{equation}}

\def\bea{\begin{eqnarray}}

\def\eea{\end{eqnarray}}

\def\bi{\begin{itemize}}

\def\ei{\end{itemize}}

\begin{document}

\title{Routes towards Anderson-like localization of Bose-Einstein condensates in disordered optical lattices}

\author{T. Schulte$^1$, S. Drenkelforth$^1$, J. Kruse$^1$, W. Ertmer$^1$, J. Arlt$^1$, K. Sacha$^2$, J. Zakrzewski$^2$, and M. Lewenstein$^{3,4,*}$}

\affiliation{$^1$ Institut f\"ur Quantenoptik, Universit\"at
Hannover, Welfengarten 1, D-30167 Hannover, Germany}

\affiliation{$^2$ Instytut Fizyki Mariana Smoluchowskiego,
Uniwersytet Jagiellonski, PL-30-059 Krakow, Poland}

\affiliation{$^3$ Institut f\"ur Theoretische Physik,
Universit\"at Hannover, D-30167 Hannover, Germany}

\affiliation{$^4$ ICFO - Institut de Ci\'encies Fot\'oniques,
08034 Barcelona, Spain}

\begin{abstract}
We investigate, both experimentally and theoretically, possible
routes towards Anderson-like localization of Bose-Einstein
condensates in disordered potentials. The dependence of this  quantum
interference effect on the nonlinear interactions and the shape of
the disorder potential is investigated. Experiments with an optical
lattice and a superimposed disordered potential reveal the lack of
Anderson localization. A theoretical analysis shows that this absence
is due to the large length scale of the disorder potential as well as
its screening by the nonlinear interactions. Further analysis shows
that incommensurable superlattices should allow for the observation
of the cross-over from the nonlinear screening regime to the Anderson
localized case within realistic experimental parameters.
\end{abstract}

\maketitle

Disordered systems have played a central role in condensed matter
physics in the last 50 years. Recently, it was proposed that
ultracold atomic gases may serve as a laboratory for disordered
quantum systems \cite{bodzio,Roth} and allow for the experimental
investigation of various open problems in that field \cite{anna}.
Some of these problems concern strongly correlated systems
\cite{Auerbach}, the realization of Bose \cite{fisher,scalettar} or
Fermi glasses \cite{fermigl}, quantum spin glasses \cite{sachdev} and
quantum percolation \cite{vera}. This letter addresses one of the
most important issues, namely the interplay of Anderson localization
(AL) \cite{pwa} and repulsive interactions \cite{anderson}. This
interplay may lead to the creation of delocalized phases both for
fermions \cite{metalglass} and bosons \cite{scalettar}. The possible
occurrence of AL has also been investigated theoretically for weakly
interacting Bose-Einstein condensates (BEC) \cite{becdis}, and in
this case it was shown that even moderate nonlinear interaction
counteracts the localization. As a main result of this letter we show
that despite this difficulty there exists an experimentally
accessible regime where Anderson-like localization can be realized
with present day techniques.

Several methods have been proposed to produce a disordered, or
quasi-disordered potential for trapped atomic gases. They include the
use of speckle radiation \cite{dainty}, incommensurable optical
lattices \cite{superlattices}, impurity atoms in the sample
\cite{Castin} and the disorder that appears close to the surface of
atom chips \cite{chiprough}.
Recently, first experiments searching for effects of disorder in
the dynamics of weakly interacting BECs were realized
\cite{inguscio}.

In this letter we shed new light on the interplay between disorder
and interactions by studying trapped BECs under the influence of a
disordered potential and a one dimensional (1D) optical lattice (OL).
The OL creates a periodic potential and the randomness of the
disordered potential leads to AL for noninteracting particles
\cite{bodzio}. We study how the presence of interactions affects
nontrivial localization in our necessarily finite system.

Our experiments were performed with $^{87}$Rb Bose-Einstein
condensates in an elongated magnetic trap (MT) with axial and radial
frequencies of $\omega_z=2\pi\times 14$~Hz and
$\omega_\perp=2\pi\times 200$~Hz, respectively. Further details of
our experimental apparatus were described previously \cite{Luigi}.
The number of condensed atoms $N$ was varied between $1.5\cdot10^4$
and $8\cdot10^4$. The OL was provided by a retro-reflected laser beam
at $\lambda=825$~nm superimposed on the axial direction of the
magnetic trap. The depth of the OL was typically set to 6.5~$E_r$,
where the recoil energy is given by $E_r=\hbar^2 k^2/2m$. For this
configuration the peak chemical potential varied between 0.25~$E_r$
and 0.5~$E_r$. The disorder potential (DP) was produced by projecting
the image of a randomly structured chrome substrate onto the atoms
giving rise to a spatially varying dipole potential along the axial
direction of the cloud. Due to the resolution of the imaging system
the minimal structure size of the DP was limited to 7~$\mu$m. We
define the depth of the DP as twice the standard deviation of the
dipole potential, analogously to \cite{inguscio}. The combined
potential allowed for the first realization of an ultracold
disordered lattice gas.

\begin{figure}[h]

\centering

\includegraphics*[width=8.6cm]{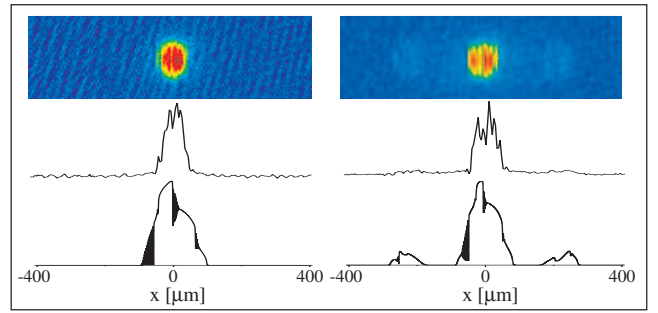}

\caption{Typical absorption images of a BEC with $N=7\cdot 10^4$
released from the combined MT plus DP (left column) and MT plus OL
plus DP (right column). The second row shows the column density and
the third row shows the result of a 1D simulation. The lattice depth
was 6.5~$E_r$ and the DP had a depth of 0.2~$E_r$.}

\label{expimg}

\end{figure}

After the production of the BEC in the MT, we performed the
following experimental sequence: We first ramped up the OL
potential over 60~ms, then the DP was ramped up over another
60~ms, followed by a hold time of 20~ms. Finally all potentials
were switched off and the atomic density distribution was measured
after 20~ms of ballistic expansion using absorption imaging.
Alternatively we performed the same experiment without the OL.

\begin{figure}

\centering

\includegraphics*[width=8.6cm]{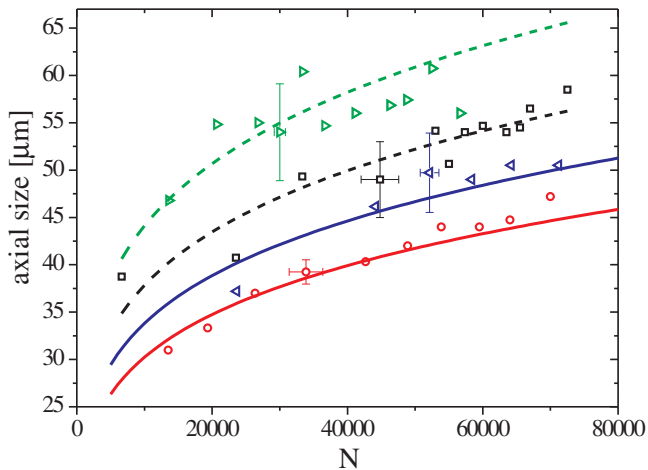}

\caption{Size of the central peak after 20~ms of ballistic expansion
versus the number of atoms. The clouds were released from the
following potentials MT (red \textcolor{red}{$\circ$}), MT plus DP
(black $\Box$), MT plus OL (blue \textcolor{blue}{$\lhd$}), MT plus
DP plus OL (green \textcolor{green}{$\rhd$}). The lines correspond to
a theoretical prediction (see text). The lattice depth was 6.5~$E_r$
and the DP had a depth of 0.1~$E_r$.}

\label{expres}

\end{figure}

Figure \ref{expimg} shows typical absorption images for the case
of DP only and for the case of combined DP and OL. The obtained
density distributions show two characteristic features. On one
hand they display pronounced fringes and on the other hand the
axial size of the central peak is modified with respect to the
case without DP. We extract the axial size of the peak by fitting
the density with a parabolic distribution. The resulting sizes are
shown as a function of the atom number in Fig.~\ref{expres}.

Both features can be attributed to the distribution of the atoms into
the wells of the DP. This can lead to a slight fragmentation of the
BEC and causes strong fringes in the absorption images. These results
are in good qualitative agreement with a numerical simulation based
on a 1D Gross-Pitaevskii equation (GPE) as shown in
Fig.~\ref{expimg}. Note that the pronounced interference fringes in
the simulation result from the interaction dominated axial expansion
within our 1D model. The additional axial confinement due to the DP
also leads to an increase of the axial size after expansion shown in
Fig.~\ref{expres}. The red and blue curves show a theoretical
prediction based on the Thomas-Fermi (TF) approximation. For the
black and green lines the same functional dependence was fitted to
the experimental data. This revealed an increase in axial size by
25\% and 28\% respectively. We have used a 3D simulation to confirm
that this increase is consistent with the modification of the
chemical potential, introduced by the DP. Note, that the change in
size depends strongly on the exact realization of the disorder.
Despite these effects of the DP, the computed ground states reveal
the absence of exponentially localized states (see analysis below)
and we conclude that the observed localization in the absorption
images is not caused by quantum interference effects, i.e. it does
not represent AL.


In order to understand the experimental results, we consider an
effective 1D model. The BEC spreads over more than a hundred wells of
the OL, each of the wells containing several hundreds of atoms. In
this situation and for depths of the OL and DP studied here the GPE
is appropriate \cite{foot1}. In oscillator units corresponding to the
trap frequency the GPE reads

\be
i\partial_t\phi=\left[-\frac{\partial^2_x}{2}+\frac{x^2}{2}+V_0\cos^2(kx)+V_{\rm
dis}(x)+g|\phi|^2\right]\phi, \label{gpe}
\ee

\noindent where $V_0$ is the depth of the OL while the DP is
represented by $V_{\rm dis}(x)$. The coupling constant $g$ is chosen
such that the TF radius equals the axial radius of the 3D atomic
cloud in the experiment (for the case of $N=7\cdot 10^4$ presented in
Fig.~\ref{expimg} we obtain $g=1500$). In all further cases we have
chosen $V_0=6.5\;E_r$.

The DP in the experiments changes on a scale much larger than the
lattice spacing and the condensate healing length, $l=1/\sqrt{8\pi n
a}$, where $n$ is the condensate density and $a$ the atomic
scattering length. This suggests the applicability of the so-called
effective mass analysis \cite{effmass}. We determine the ground state
solution of the stationary GPE in the form $\phi_0(x)=\sqrt{\cal
N}f(x)u_0(x)$, where $u_0(x)$ is the Bloch function corresponding to
the ground state of the OL potential, $f(x)$ is an envelope function
and $\cal N$ is a constant chosen such that $\phi_0$ is normalized to
unity. This  leads to an effective GPE where the OL potential is
eliminated but the mass of a "particle" and the interaction strength
become modified. For the experimental parameters the effective mass
is $m^*=2.56$ and the renormalized interaction strength for $N=7\cdot
10^4$ is $g^*=2498$.

Due to the large value of $g^*$ we may use the TF approximation
and obtain the envelope function in the form
$
|f(x)|^2=[\mu^*-x^2/2-V_{\rm dis}(x)]/g^*,
$
 where $\mu^*$ is determined from the condition $\int |f(x)|^2{\rm
d}x=1$. The squared overlap of the obtained $\phi_0$ with the exact ground
state of the GPE is 0.999 which implies that the effect of the lattice
potential is reduced to a modification of the coupling constant
for the TF profile of the combined MT plus DP. Thus, similarly to the
experiments performed in the absence of an OL \cite{inguscio} we observe a
fragmentation of the BEC induced by the DP but this fragmentation does not
correspond to Anderson-like localization.

To enter the localized regime, it is therefore necessary to introduce
a disorder that changes on a length scale comparable to the lattice
spacing.
Due to the limited resolution of the DP imaging optics this poses
considerable experimental difficulties. Alternatively one may use a
pseudorandom potential obtained with the help of two, or more
additional optical lattices with incommensurable frequencies
\cite{remark}. However, even the realization of such a fine scale
disorder is not necessarily sufficient for the observation of
non-trivial localization. Indeed, for a solution $\phi_0$ of the
stationary GPE the nonlinear term $g|\phi_0(x)|^2$ may be treated as
an additional potential. When the atoms accumulate in the wells of
the random potential, the nonlinear term in the GPE effectively
smoothes the potential modulations \cite{becdis}. For typical
experimental parameters the term $g|\phi_0(x)|^2$ dominates over
$V_{\rm dis}(x)$ and consequently the randomness necessary for
localization is lost.

\begin{figure}

\centering

\includegraphics*[width=8.6cm]{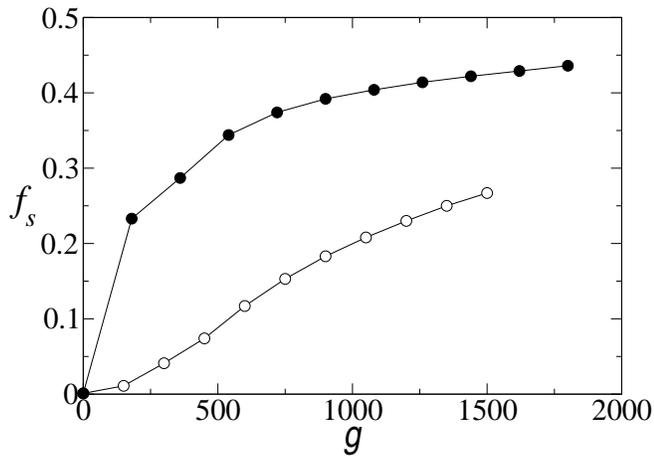}

\caption{Superfluid fraction as a function of the coupling
constant $g$ obtained from a 1D GPE simulation for a pseudorandom
potential. Full (open) symbols correspond to a
trap frequency of $2\pi\times14$~Hz ($2\pi\times4$~Hz).}

\label{theosf}

\end{figure}

This picture is confirmed by analyzing the dependence of the
superfluid fraction on the coupling constant $g$ shown in
Fig.~\ref{theosf}. To calculate the superfluid fraction we have
numerically solved the 1D GPE
in a box with periodic boundary conditions in the presence of an OL and a
pseudorandom potential created by two additional optical lattices at 960~nm and
1060~nm with depths of 0.2~$E_r$. The size of the box was chosen to match the
size of the atomic cloud in the harmonic potential. The superfluid fraction is
defined as $f_s=2[E_0(v)-E_0(0)]/Nv^2$ where $E_0(v)$ is the ground state
energy when a velocity field $v$ is imposed on the system (i.e. we compute the
ground state solution in the form $\phi_0(x)\exp(ivx)$ where $\phi_0(x)$
fulfills periodic boundary conditions) \cite{superfluid}. The superfluid
fraction remains large for typical experimental parameters, indicating the
absence of Anderson-like localization.

\begin{figure}

\centering

\includegraphics*[width=8.6cm]{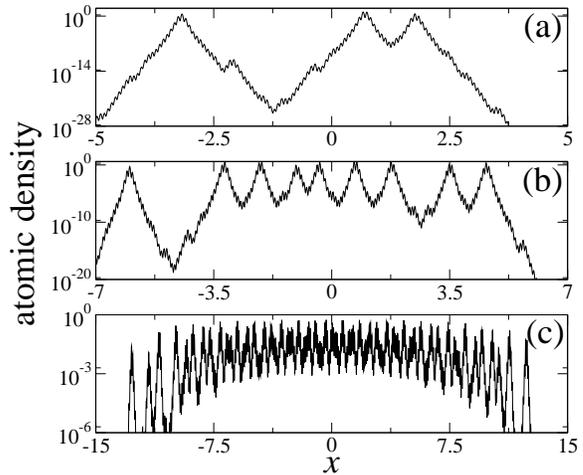}

\caption{Ground states of the GPE (note the varying logarithmic
scales) for a condensate in the combined potential of the MT, OL and
pseudorandom potential. The depth of the OL is 6.5$\;E_r$ while the
depths of the additional lattices forming the pseudorandom potential
are $0.2\;E_r$. The coupling constants $g$ for the panels are: 0.5
(a), 8 (b), 256 (c). Oscillator units corresponding to a trap
frequency of $2\pi\times 4$~Hz are used.}

\label{B}

\end{figure}

To overcome the screening of the disorder potential the interaction
within the atomic sample has to be reduced. This can be achieved by
reducing the number of atoms, lowering the trap frequencies or tuning
the scattering length via Feshbach resonances. We have performed
calculations for a trap frequency of $2\pi\times 4$~Hz and a
pseudorandom potential equivalent to the one used for
Fig.~\ref{theosf}. For $g=0$ one obtains Anderson-like localization
of the ground state wavefunction which is characterized by an
exponential localization $|\phi_0(x)|^2\propto \exp(-|x-x_0|/l)$,
with the localization length $l\approx 0.027$. For such a
non-interacting system, there might exist several localized single
particle states with an energy close to the ground state. For finite
observation times, condensation could occur into several of these low
energy states, and several "small" condensates with different
condensate wavefunctions could coexist. Figure~\ref{B} suggests that
the condensate wavefunction becomes a combination of these localized
states due to nonlinear interactions.

Increasing $g$ causes the ground state to contain a larger number of
localization centers. However, the localization length in these cases hardly
deviates from the noninteracting case. When $g$ is of the order of 500 one can
no longer distinguish individual localized states and the clear signature of
non-trivial localization vanishes. This is consistent with the appearance of a
significant superfluid fraction in Fig.~\ref{theosf}. The results shown in
Fig.~\ref{B}c for $g=256$ correspond to axial and radial frequencies of
$2\pi\times 4$~Hz and $2\pi\times 40$~Hz, respectively and $N=10^4$. In this
case the simulation shows characteristic features of Anderson-like localization
while these parameters are within experimental reach. The scenario of a
cross-over from the Anderson to the screening regime, presented here, is one of
the most important results of our analysis.

\begin{figure}

\centering

\includegraphics*[width=8.6cm]{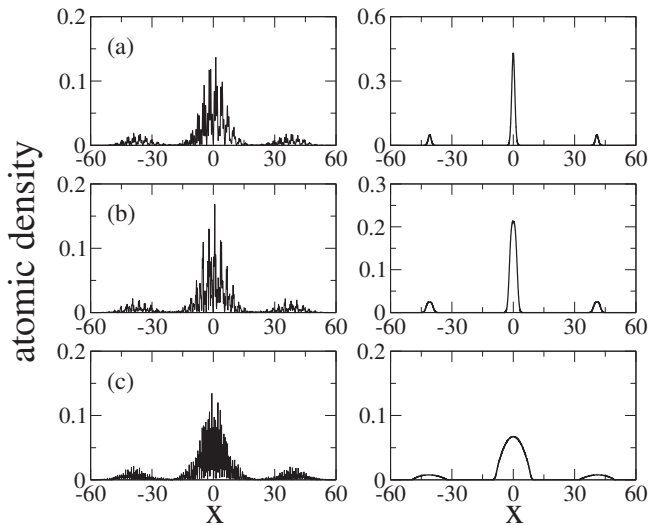}

\caption{Atomic density after 20~ms of ballistic expansion for a
condensate prepared initially in the states shown in Fig.~\ref{B}
(left column) and without DP (right column). Oscillator units
corresponding to a trap frequency of $2\pi\times 4$~Hz are used.}

\label{C}

\end{figure}

Our theoretical investigation also shows that the detection of the onset of
Anderson-like localization using a measurement of the density distribution
after ballistic expansion might be difficult. We have calculated the atomic
density profiles after 20~ms of ballistic expansion corresponding to the
parameters of Fig.~\ref{B}. Despite a striking difference in the ground state
wavefunction, the width of the envelope of the zero-momentum peak, which is
related to the localization length $l$, does not vary significantly as shown in
Fig.~\ref{C}. In addition Fig.~\ref{C}c shows that the expansion is dominated
by the interaction for experimentally accessible values of $g$ within our 1D
model. Future experiments on the detection of localization may rather rely on a
measurement of the superfluid fraction (see \cite{keith}) in an accelerated
optical lattice.

In conclusion, we have presented a detailed analysis of non-trivial
localization for slowly varying potentials and in pseudorandom
potentials in the presence of interactions. We have shown the absence
of localization in the experimental case and explained this effect
using an effective mass approach. For a truly random potential a
suppression of Anderson-like localization due to the screening by
nonlinear interactions was found. An analysis for small interactions
and a pseudorandom potential reveals the characteristic features of
Anderson-like localization. The transition from the localized to the
screened delocalized regime may be detected via an analysis of the
superfluid fraction. This work paves the way towards the observation
of Anderson-like localization in an experimentally accessible regime.

\acknowledgements

We thank L. Santos, L. Sanchez-Palencia and G.V. Shlyapnikov for
fruitful discussions. We acknowledge support from the Deutsche
Forschungsgemeinschaft (SFB 407, SPP 1116, GK 282, 436 POL), the
ESP Programme QUDEDIS, the  Polish government funds
PBZ-MIN-008/P03/2003 (KS) and 1P03B08328 (2005-08) (JZ).


\begin{thebibliography}{99}
\bibitem{foot1} We have checked that the depletion of the condensate is negligible
and the GPE is therefore well justified.
\end{thebibliography}
\end{document}